\documentclass{emulateapj}

\newcommand{\Lsol}{\mbox{$L_\odot$}}
\newcommand{\Msol}{\mbox{$M_\odot$}}
\newcommand{\Vsys}{\mbox{$V_{\rm sys}$}}
\newcommand{\asec}{\mbox{$''$}}
\newcommand{\kms}{\mbox{km s$^{-1}$}}
\newcommand{\perbeam}{\mbox{beam$^{-1}$}}
\newcommand{\persquarecm}{\mbox{cm$^{-2}$}}
\newcommand{\percubiccm}{\mbox{cm$^{-3}$}}
\newcommand{\peryr}{\mbox{yr$^{-1}$}}

\newcommand{\about}{\mbox{$\sim$}}

\newcommand{\Hone}{\mbox{\ion{H}{1}}}

\newcommand{\HH}{\mbox{H$_2$}}
\newcommand{\HCOplus}{\mbox{HCO$^{+}$}}
\newcommand{\nd}{\nodata}
\newcommand{\tnm}[1]{\tablenotemark{#1}}

\newcommand{\Tb}{\mbox{$T_{\rm b}$}}
\newcommand{\signal}{\mbox{$\sigma$}}
\newcommand{\Nhh}{\mbox{$N_{\rm H_{2}}$}}

\newcommand{\citest}[1]{\citeauthor*{#1}}

\shorttitle{P-Cygni Profiles toward Arp 220 Nuclei}
\shortauthors{Sakamoto et al.}

\slugcomment{Accepted for publication in The Astrophysical Journal Letters}

\begin{document}
\title{P-Cygni Profiles of Molecular Lines toward Arp 220 Nuclei}

\author{Kazushi Sakamoto\altaffilmark{1}, 
Susanne Aalto\altaffilmark{2}, 
David J. Wilner\altaffilmark{3}, 
John H. Black\altaffilmark{2},
John E. Conway\altaffilmark{2}, \\
Francesco Costagliola\altaffilmark{2}, 
Alison B. Peck\altaffilmark{4},
Marco Spaans\altaffilmark{5}, 
Junzhi Wang\altaffilmark{6}, \\
and
Martina C. Wiedner\altaffilmark{7}
}
\altaffiltext{1}{Academia Sinica, Institute of Astronomy and Astrophysics, P.O. Box 23-141, Taipei 10617, Taiwan; 
                        \email{ksakamoto@asiaa.sinica.edu.tw}} 
\altaffiltext{2}{Department of Radio and Space Science, Chalmers University of Technology, Onsala Space Observatory, SE 439 92 Onsala, Sweden} 
\altaffiltext{3}{Harvard-Smithsonian Center for Astrophysics, 60 Garden St., Cambridge,   MA 02138, USA}
\altaffiltext{4}{Joint ALMA Observatory, Av. El Golf 40 - Piso 18, Las Condes, Santiago 7550108, Chile}
\altaffiltext{5}{Kapteyn Astronomical Institute, P.O. Box 800, 9700 AV Groningen, The Netherlands}
\altaffiltext{6}{Department of Astronomy, Nanjing University,  22 Hankou Road, Nanjing, 210093, China}
\altaffiltext{7}{Observatoire de Paris, 61, Ave. de l'Observatoire, 75014 Paris, France}

\begin{abstract} 
We report  \about100 pc (0\farcs3) resolution observations of (sub)millimeter  
\HCOplus\ and CO lines in the ultraluminous infrared galaxy Arp 220. 
The lines peak at two merger nuclei, with \HCOplus\ being more spatially concentrated than CO.
Asymmetric line profiles with blueshifted absorption and redshifted emission 
are discovered in \HCOplus(3--2) and (4--3) toward the two nuclei  
and in CO(3--2) toward one nucleus.  
We suggest that these P-Cygni profiles are due to \about100 \kms\ outward motion of molecular gas from the nuclei.
This gas is most likely outflowing from the inner regions of the two nuclear disks rotating around individual nuclei,
clearing  the shroud around the luminosity sources there.
\end{abstract}


\section{INTRODUCTION}  
\label{s.introduction}
The two merger nuclei of Arp 220, 
about 300 pc apart on the sky \citep[see][1\asec=361 pc]{Scoville98},  
generate at least 40\% of the total luminosity of the ultraluminous infrared galaxy 
\citep[$L_{\rm 8-1000\, \mu m}=10^{12.2}$\Lsol;][]{Soifer99}.
Each has a rotating gas disk of  about 100 pc extent, 
the rotation axes of which are misaligned with one another \citep{Sakamoto99}.
The central 50--80 pc of the brighter western nucleus has a bolometric luminosity of $\geq 2\times 10^{11} \Lsol$
according to sub-arcsec imaging of submillimeter continuum \citep[hereafter \citest{Sakamoto08}]{Sakamoto08}.
There is also a kpc-scale molecular disk surrounding the two nuclei \citep{Scoville97}.
IR absorption analyses \citep[e.g.,][]{Dudley97, Haas01} suggest
the dominant luminosity source of Arp 220 to be compact and deeply buried,
as expected in the two nuclear disks, rather than widely spread, as expected in the outer disk.
Both nuclear disks host starbursts with dozens of radio supernova remnants \citep{Parra07}
and may also contain a buried quasar
(\citealt{Soifer84}; \citealt{Downes07} and references therein; \citest{Sakamoto08}).

Properties of the nuclear disks, including their structure, kinematics, and gas conditions, 
must be elucidated in order to understand fully the nature and  evolution of activity in Arp 220.
In this {\it Letter} we report our high-resolution observations of Arp 220 
in (sub)millimeter molecular lines, including
its first subarcsecond resolution imaging in \HCOplus\ 
and the first detection of P-Cygni line profiles toward the two nuclei.

\section{SUBMILLIMETER ARRAY OBSERVATIONS} 
\label{s.obs}
We observed CO(3--2) and \HCOplus(4--3) simultaneously in 2007 May
and \HCOplus(3--2) in 2008 August using the most extended configuration of the Submillimeter Array 
\citep[SMA;][]{Ho04}\footnote{
The Submillimeter Array is a joint
project between the Smithsonian Astrophysical Observatory and the
Academia Sinica Institute of Astronomy and Astrophysics, and is
funded by the Smithsonian Institution and the Academia Sinica.}.
We also used the most compact configuration  in 2009 January to measure the total \HCOplus(3--2) flux.
System gain was calibrated using two nearby quasars, J1635+381 and J1613+342.
The passband over our 2 GHz-wide sidebands was calibrated 
using other brighter quasars.
The data were reduced mostly the same way as in \citest{Sakamoto08}. 
Continuum  in each sideband was collected from 
the channels outside of radial velocities between 4800 and 5900 \kms\ in
CO(3--2), \HCOplus(4--3) and (3--2), and HCN(4--3) lines.
The continuum was then subtracted in visibilities from line data.
Phase self-calibration  was made using the continuum of Arp 220.
Line data were binned to 30 \kms\ resolution and have velocities
in the radio convention with respect to the LSR.
Our spatial resolution is about 0\farcs3 or 100 pc and
our CO(3--2) data are twice as sensitive as those in \citest{Sakamoto08}.
Since CO(3--2) was placed at the band center,
our velocity coverage for \HCOplus(4--3) in the image band is limited to 5180 -- 6840 \kms.

\section{RESULTS}
\subsection{Compact \HCOplus\ and Extended CO}
\label{s.distribution}
\begin{figure}[t]
\begin{center}
\includegraphics[scale=0.7]{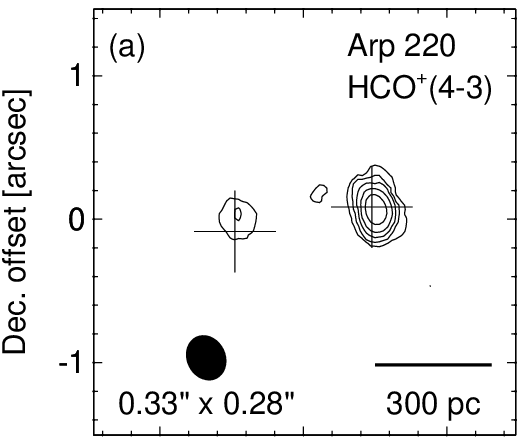} 
\includegraphics[scale=0.7]{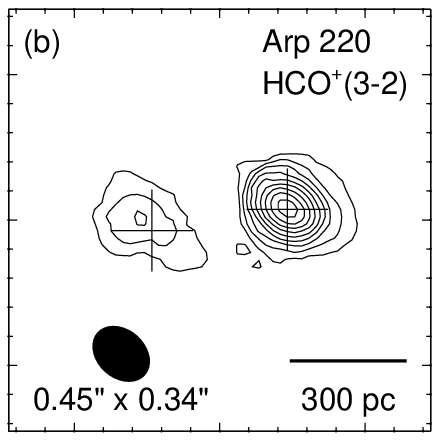} 
\\
\includegraphics[scale=0.7]{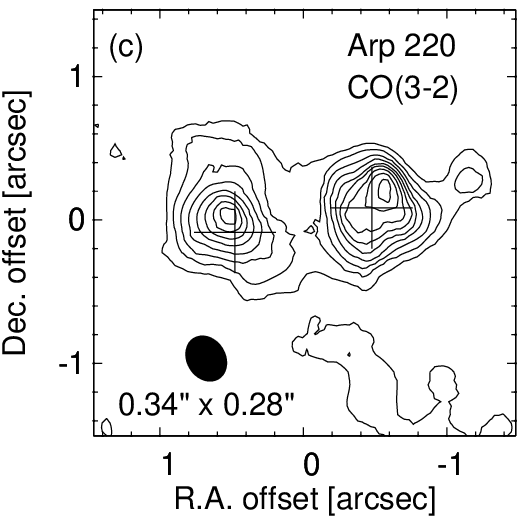} 
\includegraphics[scale=0.7]{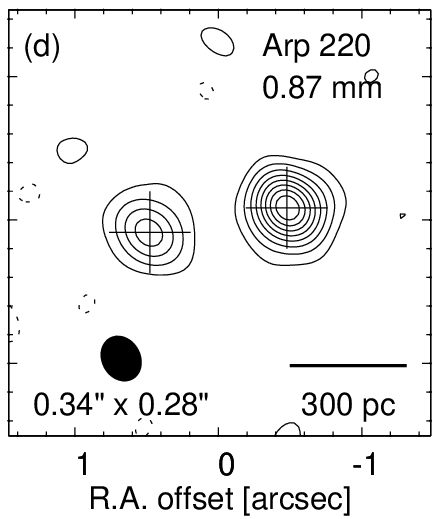} 
\end{center}
\caption{ \label{f.maps}
Integrated intensity maps of \HCOplus(4--3), (3--2), and CO(3--2) lines and 0.87 mm continuum.
The line moment-0 maps were made using positive signals only.
The lowest contour is at 2$\sigma$ and
the $n$-th contour is at $9.5n$, $4.3n$, and $10.3n$ Jy \perbeam\ \kms\ in (a), (b), and (c), respectively,
and at $14.4 n^{1.5}$ mJy \perbeam\ in (d).
Crosses are at continuum peaks and filled ellipses show synthesized beams.
}
\end{figure}

We detected \HCOplus, CO, and $\lambda$\about1 mm continuum emission associated with the two nuclei (Fig. \ref{f.maps}). 
The western nucleus is more luminous than the eastern, by a factor of 2--3 in \HCOplus\ and continuum.
The concentration of  \HCOplus\ emission to the two nuclei is real because
we detected most of the total line flux in Arp 220 (Table \ref{t.param}).
This is in contrast to CO(3--2) where only 23\% of the single-dish flux was recovered even though 
CO(3--2) and \HCOplus(4--3) had almost the same u-v coverage.
CO(3--2) emission is indeed extended on scales greater than 2\asec\ in \citest{Sakamoto08}.
Thus  \HCOplus(4--3) and (3--2) emission of Arp 220 is mostly from the two nuclear disks
while the majority of CO(3--2) comes from the outer disk around them.
Moreover, in each nuclear disk, the \HCOplus\ emitting region is more compact than the CO emitting region
judging from their half-peak sizes in Fig. \ref{f.maps}.

The peak brightness temperatures (\Tb) of the lines at or around each nucleus, listed in Table \ref{t.param},
indicate warm molecular gas in the nuclear disks.
The CO  peaks are about 50 K and \HCOplus\ about 20 K. 
These high \Tb\ and high \HCOplus-to-CO intensity ratios in the nuclear disks
imply that molecular gas exists there in conditions favorable for the transitions of high critical density.

\subsection{P-Cygni Line Profiles at Both Nuclei}
\label{s.line_profile}
\begin{figure}[t]
\epsscale{1.0}
\plottwo{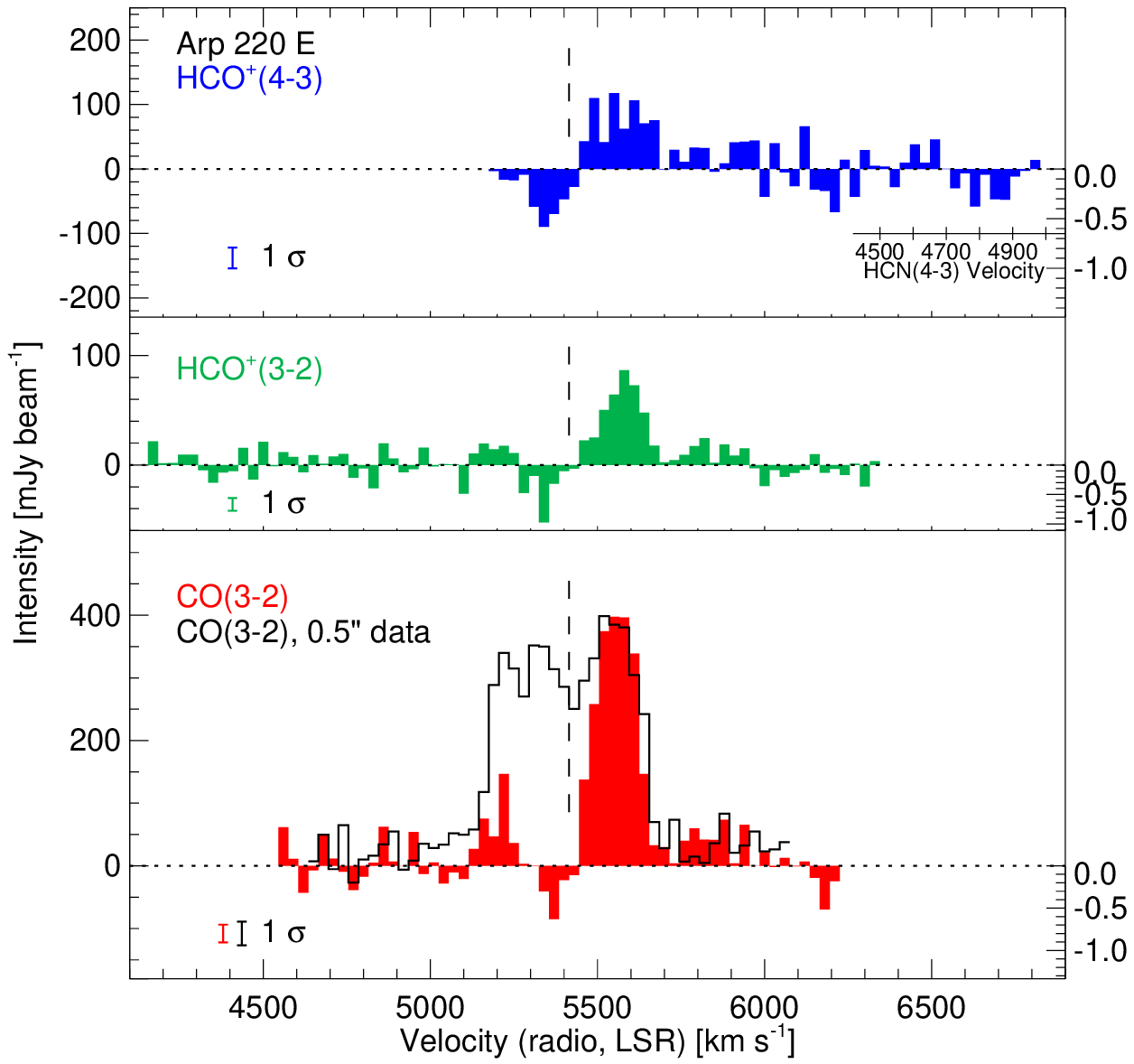}{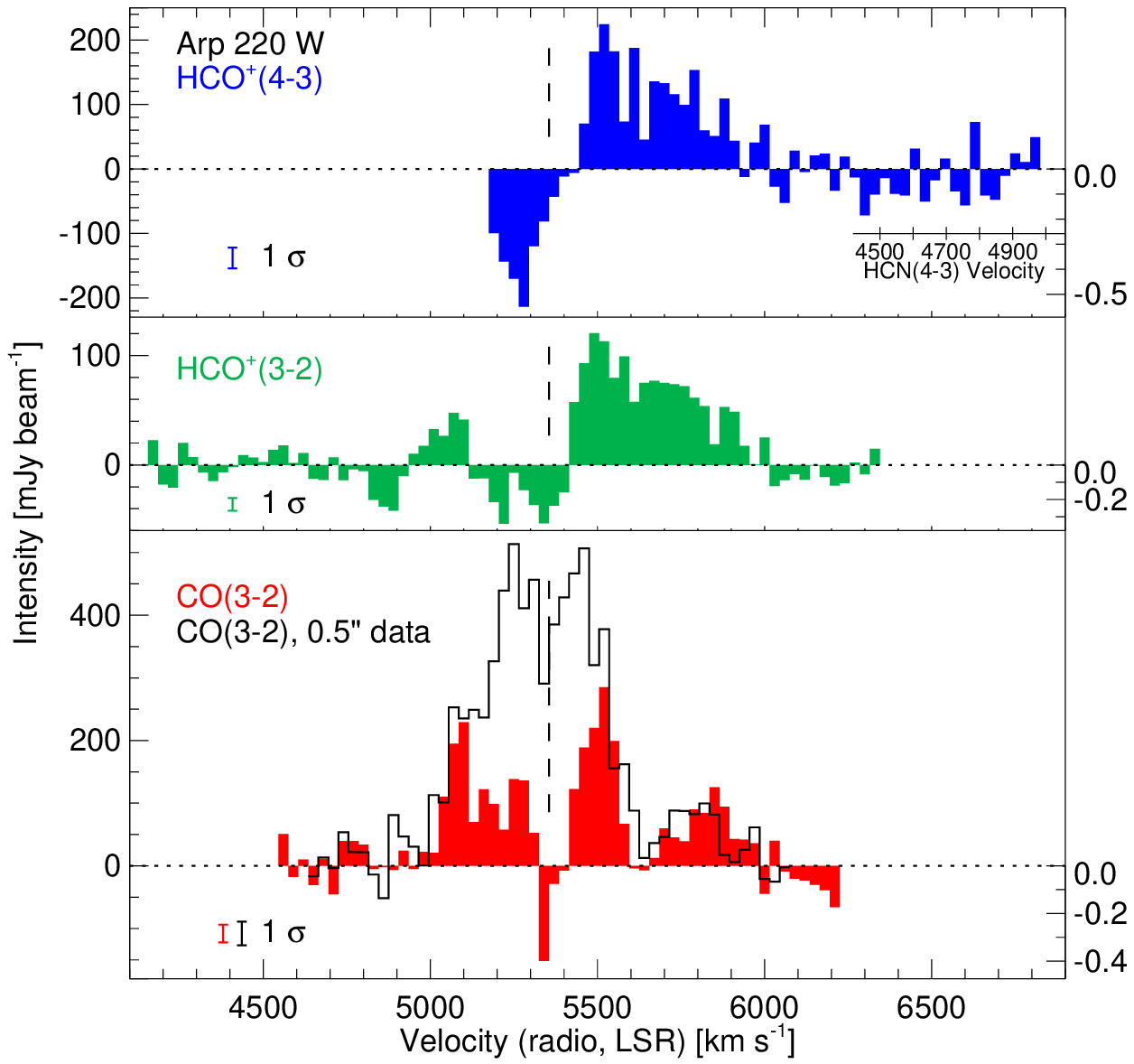}  
\caption{ \label{f.spectra}
Molecular-line spectra at the two nuclei of Arp 220. Continuum has been subtracted.
The right ordinate is the fractional absorption depth $f_{\rm a}$ 
with respect to the continuum intensity of each nucleus, and
can be converted to optical depth via  $\tau_{\rm a} = - \log(-f_{\rm a})$.
The 1\signal\ noise is 32, 12, and 28 mJy \perbeam\ for \HCOplus(4--3), (3--2), and CO(3--2),
respectively. 
The CO(3--2) spectra overplotted in black are from 0\farcs5-resolution data in \protect\citet{Sakamoto08}.
The systemic velocity that we estimated for each nucleus is shown by a dashed line.
}
\end{figure}

The line spectra at the two nuclei, in Fig. \ref{f.spectra}, 
show both emission and absorption.
In particular, all the observed lines except CO(3--2) toward the western nucleus
show blueshifted absorption and redshifted emission, i.e., P-Cygni profiles.
This feature is absent in our calibrator spectra.

The western nucleus has a narrow, deep absorption feature at 5340 \kms\ 
and emission at higher and lower velocities in our CO data.
This profile is consistent with the previous CO(2--1) profile in \citet{Downes07} except that
the absorption is deeper in our CO(3--2) data.
The deep absorption is almost at the systemic velocity of the nucleus, marked with a dashed line in Fig. \ref{f.spectra},
as expected if the CO-absorbing gas is in circular rotation in the western nuclear disk. 
The systemic velocity, $\Vsys({\rm W})=5355 \pm 15$ \kms, was estimated from
the central velocity of the CO profile in the low-resolution (0\farcs5) data 
of \citest{Sakamoto08},  measured at half-maximum.
The \HCOplus(4--3) and (3--2) lines are also in absorption at the systemic velocity.
In addition, both \HCOplus\ lines have absorption at blueshifted velocities
while most of their emission appears at redshifted velocities.
The \HCOplus(3--2) profile shows multiple velocity components in absorption; 
the minima at 4890, 5220, and 5340 \kms\ 
are at 3.6\signal\, 4.6\signal, and 4.6\signal, respectively. 
Absorption deeper than 3\signal\ is seen down to $\Vsys({\rm W}) - 495$ \kms.

The eastern nucleus has very asymmetric line profiles 
with respect to its systemic velocity, which
we estimate to be $\Vsys({\rm E})  =  5415 \pm15$ \kms\ 
in the same way as for the western nucleus.
In all the observed lines, 
absorption is predominantly seen at or blueward of the systemic velocity 
while most emission is in redshifted velocities, making the profiles P Cyg type.
The deepest absorption is at 45--75 \kms\ below \Vsys(E) and is
2.8\signal, 4.5\signal, and 3.0\signal\ in \HCOplus(4--3), (3--2), and CO(3--2), respectively.

\subsection{Absorption Depth}
\label{s.abs_depth}
Absorption (i.e., negative intensity) in our continuum-subtracted data
means absorption of the continuum, because line self-absorption alone cannot 
produce {\it negative} intensity.
In order to absorb continuum the excitation temperature of the foreground gas must be below the
brightness temperature of the background continuum. 
The (deconvolved) western nucleus has \Tb= 90--160  K and a size of 50--80 pc
in 860 \micron\ continuum
(\citest{Sakamoto08}; for the SED of the two nuclei see \citealt{Matsushita09}). 
The high \Tb, due to high dust opacity in submillimeter, makes line absorption 
against dust continuum more likely to occur (and deeper) than at longer wavelengths.

The maximum apparent 
optical depth of each absorption line is in the range of 0.4--1.0 
except the \HCOplus(3--2) optical depth of 3.7 (or $>$1.4 when allowing for 1 $\sigma$ error) 
toward the eastern nucleus.
The total line absorption $\int \! \tau_{\rm a} \, dV$ integrated over the velocities where $\tau_{\rm a} \ge 0.1$ 
is, in the order of \HCOplus(4--3), \HCOplus(3--2), and CO(3--2), 
75, 150, and 50 \kms\ toward the eastern nucleus and
85, 82, and 15 \kms\ toward the western.
The absorption profiles vary among the lines, more notably toward the western nucleus.
There are two caveats to keep in mind.
First, our observations provide lower limits to total absorption 
because emission from the absorbing gas itself and from ambient (i.e., non-absorbing) gas in our beam 
masks the absorption.
Second, our synthesized beam is larger for \HCOplus(3--2) than for \HCOplus(4--3) and CO(3--2).

The absorbing column  density $\Nhh$ is on the order of $10^{22}$ \persquarecm\  for the CO
absorption and $10^{23}$ \persquarecm\  for the \HCOplus\ if the absorbing gas is in 50 K LTE
and abundances are [CO/\HH]  $\approx 10^{-4}$ 
and [\HCOplus/\HH] $\approx 10^{-8}$.
The high abundance adopted for \HCOplus\ is consistent with recent PDR and XDR
models \citep{Meijerink05}, as well as models of \HCOplus\ enhancement
in protostellar outflow shocks \citep{Rawlings04}.
The absorbing gas will have a depth on the order of 1--10 pc if its density is at the CO(3--2) critical density
of $10^4$ \percubiccm.  
The actual absorbing material can be distributed over a longer length if it is inhomogeneous as in
numerical hydrodynamical simulations of multi-phase gas obscuring active nuclei \citep[e.g.,][]{Wada02}. 
The column densities of the absorbing gas calculated here have large uncertainties because the 
absorption data provide lower limits, gas excitation conditions are poorly known, 
and molecular abundances can significantly vary. 
Still, it is possible for the absorbing column density to take these values
because they are smaller than 
the total column density of the nuclei (\about$10^{25}$ \persquarecm ; \citest{Sakamoto08}).
The remarkable detection of vibrationally excited HCN in absorption in Arp 220 \citep{Salter08} suggests 
that the excitation of \HCOplus\ may also be more unusual than described by our simple analysis. 
More realistic models of the excitation and radiative transfer of both molecules, 
including radiative coupling to the intense infrared continuum, will be explored in a future publication.

\subsection{Gas Motion --- Rotation and Outflow}
\label{s.dynamics}
\begin{figure}[t]
\begin{center}
\includegraphics[scale=0.6]{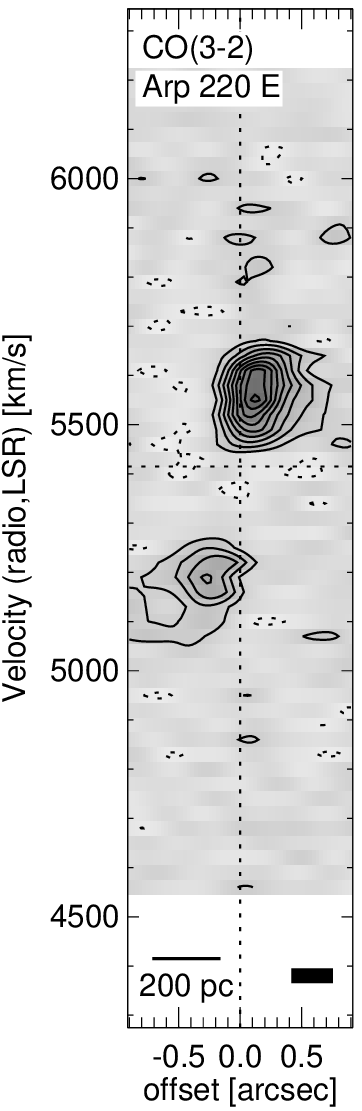} 
\includegraphics[scale=0.6]{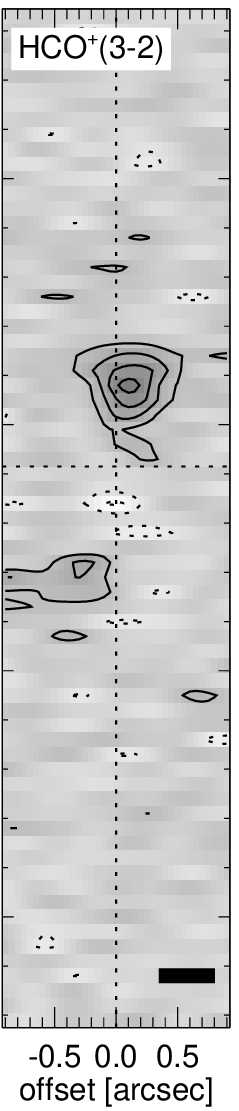} 
\includegraphics[scale=0.6]{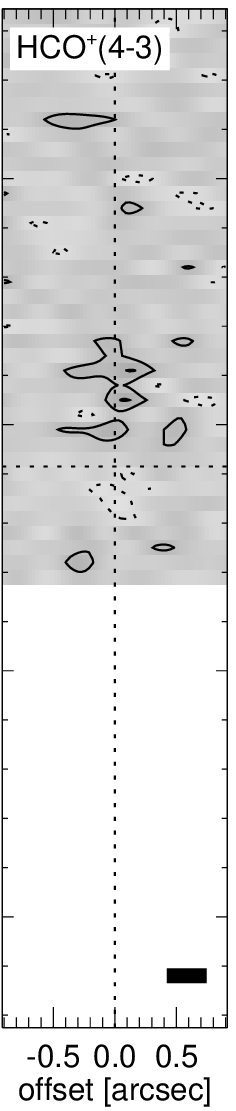}
\hspace{1mm}
\includegraphics[scale=0.6]{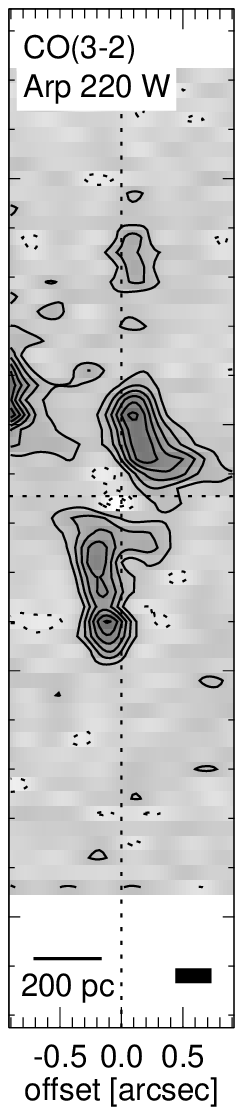} 
\includegraphics[scale=0.6]{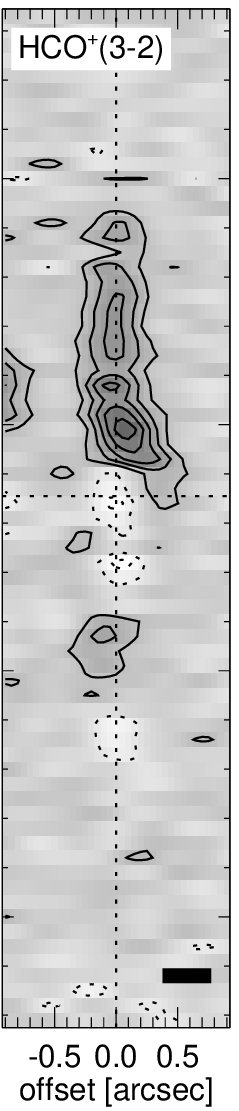} 
\includegraphics[scale=0.6]{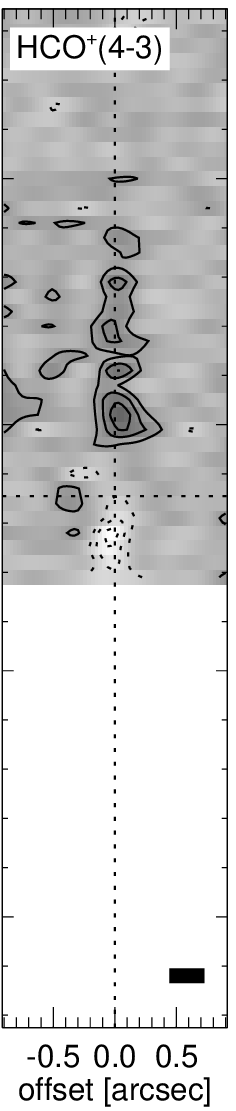} 
\end{center}
\caption{ \label{f.pv}
Position-velocity diagrams 
of Arp 220 nuclei. 
Position is measured from each nucleus along the position angle of 45\arcdeg\ for the eastern nucleus
and 270\arcdeg\ for the western.
Contours are at $2n \sigma$ $(n=\pm1, \pm2, ...)$  with negative ones dashed.
The rms noise in brightness temperature is 3.1, 1.3, and 3.5 K for CO(3--2), \HCOplus(3--2), and (4--3), 
respectively.
The horizontal dotted lines are at the systemic velocities of the nuclei, and
the black rectangles show the spatial and velocity resolutions.}
\end{figure}

Rotation of gas around each nucleus is evident in the CO(3--2) position-velocity diagrams (Fig. \ref{f.pv})
as the negative to positive shift of $V- \Vsys$ across the nucleus. 
The CO emission ranges from 5000 to 5900 \kms\ around the western nucleus.
The high-velocity gas around 5800--5900 \kms\ is detected at $>$3 \signal\ for the first time.
The fall-off of the rotation velocity away from the nucleus agrees with the CO(2--1) observations by \citet{Downes07}
and suggests truncation of the mass distribution of the nucleus beyond a $\lesssim$100 pc core.

Radial motion of gas, in addition to the rotation, is suggested by the blueshifted absorption mentioned above.
The absorbing gas must be in front of the continuum core of each nucleus and moving away from it toward us. 
Although absorption only tells us the gas motion along our lines of sight to the continuum nuclei, 
the detection of similar blueshifted absorption on both nuclei whose nuclear disks are misaligned suggests that
the gas motion is more likely a radial outflow in most directions 
than non-circular motion only along our sight lines. 
The redshifted emission in the P-Cygni profiles must then contain emission from gas 
on the far side of the continuum nuclei and moving away from the nuclei and us. 
The other emission in the spectra must be from the gas rotating in the nuclear disks.

The typical outflow velocity of the absorbing gas is probably about 100 \kms\ because 
the deepest absorption in the P-Cygni spectra 
is between $(\Vsys -45)$ and $(\Vsys - 135)$ \kms.
High-velocity components up to \about500 \kms\ are suggested toward the western nucleus
by the \HCOplus(3--2) absorption with large velocity offsets.
The absorbing gas is turbulent or has a velocity gradient across each continuum nucleus
because absorption is also seen at slightly positive velocities
with respect to systemic.
The mass of the outflowing gas and the mass outflow rate would be
on the order of $5\times 10^{7}$ \Msol\ and 100 \Msol\ \peryr\ if the outflowing
gas with a column density of $10^{23}$ H$_2$ \persquarecm\  is mostly within a radius of 50 pc 
(see \S \ref{s.loc_absorber}) and has an isotropic outward velocity of 100 \kms.
Note however that these numbers strongly depend on the geometry of the gas flow.

\section{DISCUSSION}
\subsection{Location of the Absorber}
\label{s.loc_absorber}
The blueshifted absorbing gas is probably located in the two nuclear disks.
This is likely because \HCOplus\ {\it emission} is concentrated to the two nuclear disks.
The nuclear disks thus have significant column densities of \HCOplus\ excited to the J=3 and 4 levels,
whether the excitation is collisional (with critical densities of $10^{6}$--$10^{7}$ \percubiccm\ at 50 K)
or by radiation.
The J=3 \HCOplus\ in front of a bright source causes \HCOplus(4$\leftarrow$3) absorption.
Since column densities of excited \HCOplus\  must be less outside the nuclear disks according to our
observations,  any external gas is unlikely to be the dominant absorber.

In our preferred model, 
each nuclear gas disk has an inner region with significant outflow motion 
and an outer region dominated by rotation.
This configuration can partly explain 
why the P-Cygni profiles are more prominent in \HCOplus\ than in CO.
Excited \HCOplus\ is more concentrated toward the outflowing disk centers, while for extended CO
more emission from the rotating (i.e., non-outflowing) disk likely masks absorption by the outflowing gas.
It is also likely that the lower excitation temperature of \HCOplus\ expected from 
its larger Einstein A coefficients makes it a better absorber than the CO at the same location with
a higher excitation temperature.
The rotation-dominated part of the nuclear disk is likely the site of possible weak HNC maser
that \citet{Aalto09} suggested for the western nucleus
on the basis of a narrow and bright spectral feature at the systemic velocity.

A less likely location of the absorbing gas is a few 100 pc or more from the two nuclear disks.
The gas may be a part of the merger's large-scale ($>$ 10 kpc) superwind 
perpendicular to the kpc-scale outer molecular disk  \citep{Heckman87,McDowell03}.
The 5340 \kms\ absorption toward both nuclei could be explained by this common envelope model. 
However, the total mass of such a gas structure would be quite large; \about$10^{9}$ \Msol\
for an expanding envelope of 500 pc radius and a column density of  $\Nhh \sim 10^{22.5}$ \persquarecm. 
Such an expanding molecular envelope has not been detected in emission.

\subsection{Relation to Previous Observations} 
\label{s.prev_obs}
Observational signs of outflows from individual nuclei in Arp 220 have been scarce
despite ample observations of the merger's superwind at larger scales.
A biconical distribution of OH maser around the western nucleus along its rotation axis (i.e., N-S direction)
was observed by \citet[their Fig. 4]{Rovilos03}.
\citet{Aalto09} suggested that this emission as well as the faint HNC emission they found to be extended in
the same direction may be due to an outflow from the nucleus.
While kinematical evidence for the outflow is absent or inconclusive in
these subarcsecond resolution observations, \citet{Baan89} detected
highly blueshifted OH maser emission with a \about 2\arcmin\ beam and 
attributed it to an outflow at 10--100 pc from the luminosity source with
wind acceleration modeling.
In \Hone\ observations by \citet{Mundell01}, velocities of the main absorption 
toward the nuclei are $V$(E, main \Hone) = 5396 \kms\ and $V$(W, main \Hone) = 5325 \kms\
after conversion from optical-heliocentric to radio-LSR.
They are marginally  blueshifted, by 19 and 30 \kms, from our CO-based systemic velocities.
Also, the \Hone\ absorption toward the western nucleus has a subcomponent at about $-$80 \kms\
from the main component. 
These may be related to the molecular outflows that we suggest.

\subsection{Nuclear Winds and the Nuclear Disks}
\label{s.properties}
The P-Cygni profiles toward the two nuclei provide a new clue
to the energetic nucleus of Arp 220. 
The nuclear disks need to be reinvestigated in light of the suggested nuclear winds.

The compact and luminous sources in the nuclei 
probably make the outflows inevitable.
For example, supernovae inject $3\times 10^{50}$ J of mechanical energy into the western nucleus in 1 Myr 
for the supernova rate of \about3 \peryr\ that \citet{Lonsdale06} observed. 
(The rate is \about1 \peryr\ for the eastern nucleus.)
If 10\% of this is passed to kinetic energy of $10^8$ \Msol\ of gas, the gas velocity will
be 500 \kms. 
Gas at shallow parts of the gravitational potential or getting more energy 
than average will escape the nucleus. 
Although the outflow may be mostly ionized, it will entrain molecular gas on its way. 
Radiation pressure on dust also drives an outflow in the case of a very compact starburst or an AGN buried in dust
\citep{Scoville01,Scoville03,Murray05}.
The momentum of a $5\times 10^{7}$ \Msol\ gas outflowing at 100 \kms\ is equivalent to the total momentum of photons
that a $0.5\times 10^{12}$ \Lsol\ source radiates in 0.5 Myr.
Such an outflow may be more isotropic near its origin than the outflow caused by a disk starburst
and would be more significant in the inner region of each nuclear disk 
because the radiation pressure declines with radius as $r^{-2}$.

The winds are important for the evolution of the nuclear disks
because they remove the shroud of and fuel for the buried energy sources.
The small crossing time of the nuclear winds over each disk,  $\lesssim$1 Myr, is suggestive in
terms of the evolution.
It is either that we are seeing very young winds
or that the gas motion in each disk is not purely outward in every direction
despite the large opening angles of the outflows inferred in \S \ref{s.dynamics}.
For the latter, it is conceivable that high-density gas in the mid-plane of each nuclear disk is 
little affected by the outflow 
and even has inward motion due to viscosity to replenish gas to the center, as in \citet{Wada02}. 
The evolution of the nuclear gas disks is then determined by the balance of
the gas accretion, gas outflow, and gas consumption by star formation and AGN
until the coalescence of the nuclei and of the gas disks.

\acknowledgements
We are grateful to the SMA staff who made these observations possible.
We thank the referee for helpful comments and
K. S. thanks Dr. Keiichi Wada for stimulating discussion.
This research made use of 
the NASA/IPAC Extragalactic Database (NED) and
NASA's Astrophysics Data System (ADS).

{\it Facilities:} \facility{SMA}


\clearpage

\begin{deluxetable}{lrrrc}
\tablewidth{0pt}
\tablecaption{Observed Parameters  of Arp 220 \label{t.param} }
\tablehead{ 
	\colhead{Parameter}  &
	\colhead{East} &	
	\colhead{West} &
	\colhead{All} &
	\colhead{$f_{\rm tot}$\tnm{1}} 
}
\startdata
$S_{\rm CO(3-2)}$  [Jy \kms] & 250  & 301 & $(7.0 \pm 1.1)\times10^2$ & $0.23\pm0.06$ \\
$S_{\rm HCO^{+}(4-3)}$  [Jy \kms] & 19  & 61  & $(1.1\pm 0.2)\times10^2$ & $1.0\pm0.3$ \\
$S_{\rm HCO^{+}(3-2)}$  [Jy \kms] & 35  & 63 & $(1.0\pm 0.1) \times10^2$ & $0.64\pm 0.13$ \\
$S_{\rm 0.87\, mm}$  [Jy] & 0.18 & 0.49 &  $0.68 \pm 0.10$ & $1.0\pm0.2$ \\ 
$S_{\rm 1.1\, mm}$  [Jy] & 0.07 & 0.20 &  $0.27\pm 0.03$ & $0.82\pm0.15$  \\
CO(3--2) $\max \Tb$ [K] & 58 (66) & 40 (47) & \nd & \nd  \\
\HCOplus(4--3) $\max \Tb$ [K] & 16 (24) & 26 (34) & \nd & \nd \\
\HCOplus(3--2) $\max \Tb$ [K] & 12 (17) & 15 (21) & \nd & \nd  \\
0.87 mm $\max \Tb$ [K] &  16 (23) &  42 (50) & \nd & \nd  \\
1.1 mm $\max \Tb$ [K] & 7 (12) & 20 (25) & \nd & \nd \\
$V_{\rm sys}$(radio, LSR) [\kms] & 5415$\pm$15 & 5355$\pm$15 & \nd & \nd
\enddata
\tablecomments{
We conservatively estimate flux-calibration 1$\sigma$ errors to be 10\% and 15\% for 0.86 and 1.1 mm bands, respectively.
Line fluxes are integrated over 4800--5900 \kms\ [5200--5900 \kms\ for \HCOplus(3--2)] including both emission and absorption.
Peak brightness temperatures are measured at or around each
nucleus without correcting for beam dilution or missing flux. 
The first one is Rayleigh-Jeans \Tb\ and the second in parenthesis is Planck \Tb.
}
\tablenotetext{1}{
Fraction of the flux  (or flux density) recovered in our 0\farcs3-resolution data 
with respect to the total flux of Arp 220.
The total fluxes are 
$106\pm23$ Jy \kms\ for \HCOplus(4--3) \protect\citep{Greve09}, 
$(3.0\pm0.6)\times 10^{3}$ Jy \kms\ for CO(3--2) in the central 14\asec\ \protect\citep{Wiedner02}, 
$152\pm26$ Jy \kms\ for \HCOplus(3--2) (This work. SMA sub-compact array observations
with the minimum projected baseline of 9 m),
$0.7 \pm 0.1$ Jy for 0.87 mm continuum (\protect\citest{Sakamoto08} and reference therein),
and 
$0.33 \pm 0.05$ Jy for 1.1 mm continuum (This work, SMA sub-compact configuration). }
\end{deluxetable}


\begin{thebibliography}{}	
\bibitem[Aalto et al.(2009)A09]{Aalto09}
	Aalto, S., Wilner, D., Spaans, M., Wiedner, M., Sakamoto, K.,
	Black, J., \& Caldas, M.
	2009, \aap, 493, 481
\bibitem[Baan et al.(1989)]{Baan89}
	Baan, W. A., Haschick, A. D., \& Henkel, C.
	1989, \apj, 346, 680	
\bibitem[Downes \& Eckart(2007)]{Downes07}
	Downes, D., \& Eckart, A.
	2007, \aap, 468, L57 
\bibitem[Dudley \& Wynn-Williams(1997)]{Dudley97}
	Dudley, C. C., \& Wynn-Williams, C. G.
	1997, \apj, 488, 720		
\bibitem[Greve et al.(2009)]{Greve09}
	Greve, T. R., Papadopoulos, P. P., Gao, Y., \& Radford, S. J. E.
	2009, \apj, 692, 1432
\bibitem[Haas et al.(2001)]{Haas01}
	Haas, M., Klaas, U., M\"{u}ller, S. A. H., Chini, R., \& Coulson, I.
	2001, \aap, 367, L9	
\bibitem[Heckman et al.(1987)]{Heckman87}
	Heckman, T. M.,  Armus, L., \& Miley, G. K.
	1987, \aj, 93, 276 
\bibitem[Ho et al.(2004)]{Ho04}
	Ho, P. T. P., Moran, J. M., \& Lo, K. Y.
	2004, \apjl, 616, L1	
\bibitem[Lonsdale et al.(2006)]{Lonsdale06} 
	Lonsdale, C.~J., Diamond, P.~J., Thrall, H., Smith, H.~E., \& Lonsdale, C.~J.
	2006, \apj, 647, 185
\bibitem[Matsushita et al.(2009)]{Matsushita09}
	Matsushita, S. et al.
	2009, \apj, 693, 56		
\bibitem[McDowell et al.(2003)]{McDowell03}
	McDowell, J. C., et al.
	2003, \apj, 591, 154	
\bibitem[Meijerink \& Spaans(2005)]{Meijerink05}
	Meijerink, R., \& Spaans, M.
	2005, \aap, 436, 397	
\bibitem[Mundell et al.(2001)]{Mundell01}
	Mundell, C. G., Ferruit, P., \& Pedlar, A.
	2001, \apj, 560, 168	
\bibitem[Murray et al.(2005)]{Murray05}
	Murray, N., Quataert, E., and Thompson, T. A.
	2005, \apj, 618, 569	
\bibitem[Parra et al.(2007)]{Parra07}
	Parra, R., Conway, J. E., Diamond, P. J., Thrall, H., Lonsdale, C.~J., Lonsdale, C.~J., \& Smith, H.~E.
	2007, \apj, 659, 314	
\bibitem[Rawlings et al.(2004)]{Rawlings04} 
	Rawlings, J.~M.~C., Redman, M.~P., Keto, E., \& Williams, D.~A.
	2004, \mnras, 351, 1054	
\bibitem[Rovilos et al.(2003)]{Rovilos03}
	Rovilos, E., Diamond, P. J., Lonsdale, C. J., Lonsdale, C. J., \& Smith, H. E.
	2003, \mnras, 342, 373		
\bibitem[Sakamoto et al.(1999)]{Sakamoto99}
	Sakamoto, K., Scoville, N. Z., Yun, M. S., Crosas, M., Genzel, R., \& Tacconi, L. J.
	1999, \apj, 514, 68	
\bibitem[Sakamoto et al.(2008)S08]{Sakamoto08}
	Sakamoto, K. et al.
	2008, \apj, 684, 957  (S08)
\bibitem[Salter et al.(2008)]{Salter08} 
	Salter, C.~J., Ghosh, T., Catinella, B., Lebron, M., Lerner, M.~S., Minchin, R., \& Momjian, E.
	2008, \aj, 136, 389 	
\bibitem[Scoville et al.(1997)SYB97]{Scoville97}
	Scoville, N. Z., Yun, M. S., \& Bryant, P. M.
	1997, \apj, 484, 702	
\bibitem[Scoville et al.(1998)]{Scoville98}
	Scoville, N. Z.,	et al. 
	1998, \apjl, 492, L107
\bibitem[Scoville et al.(2001)]{Scoville01}
	Scoville, N. Z., Polletta, M., Ewald, S., Stolovy, S. R., Thompson, R., and Rieke, M. 
	2001, \apj, 122, 3017	
\bibitem[Scoville (2003)]{Scoville03}
	Scoville, N. Z.
	2003, J. Korean Astron. Soc., 36, 167		
\bibitem[Soifer et al.(1984)]{Soifer84}
	Soifer, B. T. et al.
	1984, \apjl, 283, L1	
\bibitem[Soifer et al.(1999)]{Soifer99}
	Soifer, B. T. et al.
	1999, \apj, 513, 207	
\bibitem[Wada \& Norman(2002)]{Wada02}
	Wada, K., \& Norman, C. A.
	2002, \apjl, 566, L21	
\bibitem[Wiedner et al.(2002)]{Wiedner02} 
	Wiedner, M.~C., Wilson, C.~D., Harrison, A., Hills, R.~E., Lay, O.~P., \& Carlstrom, J.~E.
	2002, \apj, 581, 229 	
\end{thebibliography}
\end{document}